\documentclass[mnras, twocolumn]{aastex631} 
\usepackage[T1]{fontenc}
\usepackage{graphicx}
\usepackage{amsmath}
\usepackage{color}
\usepackage{comment}
\usepackage{units}
\usepackage[normalem]{ulem}
\usepackage{hyperref}

\newcommand{\bilby}{\textsc{bilby}~}

\newcommand{\SPA}{School of Physics and Astronomy, Monash University, Clayton VIC 3800, Australia}
\newcommand{\OzGravMonash}{OzGrav: The ARC Centre of Excellence for Gravitational Wave Discovery, Clayton VIC 3800, Australia}

\begin{document}

\title{Searching for gravitational waves from compact binary mergers powering long gamma-ray bursts during LIGO-Virgo-KAGRA's O3 run}

\author{Mallika R. Sinha}
\email{mallika.sinha@monash.edu}
\affiliation{\SPA}
\affiliation{\OzGravMonash}

\author{Teagan A.~Clarke}
\affiliation{Department of Physics, Princeton University, Princeton, New Jersey, 08544, USA}
\affiliation{\SPA}
\affiliation{\OzGravMonash}

\author{Qifang Zhang}
\affiliation{\SPA}
\affiliation{\OzGravMonash}

\author{Nikhil Sarin}
\affiliation{Kavli Institute for Cosmology, University of Cambridge, Madingley Road, CB3 0HA, UK}
\affiliation{Institute of Astronomy, University of Cambridge, Madingley Road, CB3 0HA, UK}

\author{Eric Thrane}
\affiliation{\SPA}
\affiliation{\OzGravMonash}

\author{Paul D. Lasky}
\affiliation{\SPA}
\affiliation{\OzGravMonash}

\begin{abstract}
Neutron star binary mergers are often associated with short gamma-ray bursts (GRBs), but the recent detection of kilonovae coincident with long GRBs suggest that some mergers may produce long GRBs. Motivated by these developments, we perform a search for binary neutron star and neutron star-black hole gravitational-wave signals coincident with long GRBs using data from the third LIGO--Virgo--KAGRA (LVK) observing run. We analyze LVK data coincident with long GRBs detected by Fermi's GRB Monitor and Swift's Burst Alert Telescope when at least two gravitational-wave observatories were running. We find no evidence of a coincident gravitational-wave signal and set limits on the luminosity distance to each of these long GRBs under the assumption that they were powered by binary mergers.

\end{abstract}

\section{Introduction}
Gamma-ray bursts (GRBs) are extremely energetic emissions of gamma rays that can last anywhere from milliseconds to hours~\citep[e.g.,][]{Zhang_book, Neights_2025}. The progenitors of GRBs have been debated since their discovery. Bimodality in their temporal and spectral properties has given rise to the classification of short-hard ($<2$\,s) and long-soft ($>2$\,s) GRBs~\citep{Kouveliotou_1993}. The former are thought to be produced by mergers of a neutron star with another neutron star or stellar-mass black hole~\citep{Paczynski_1991, Eichler_1989, Popham_1999, Narayan_1992} (supported by the coincident GRB and gravitational-wave detection of GW170817 and GRB170817A~\citep{Abbott_2017_170817,Abbott_2017_grb, savchenko_2017, goldstein_2017}), and the latter from the core collapse of massive stars~\citep{Woosley_1993, Paczy_ski_1998, Popham_1999}.

Recent observations of three long GRBs casts doubt on the above classification, with evidence their progenitors were compact binary mergers involving at least one neutron star. Optical/infrared kilonovae signals have been associated with GRBs GRB060614~\citep{gehrel_2006, della_2006, Zhang_2007, Jin_2015}, GRB211211A~\citep{Rastinejad_2022, troja_2022, Yang_2022, zhang_2022} and GRB230307A~\citep{yi2025,Levan_2023, sun_2024, yang_2023, Maccary_2026, gillanders2025analysisjwstspectrakilonova}, suggesting the duration of GRBs may not uniquely map to the type of progenitor.\footnote{Some suggested influential factors other than the progenitor include energy dissipation processes, environmental factors, disk mass, and the emission mechanism~\citep[e.g.,][]{yi2025, gottlieb2023, Gompertz_2022, zhu2025, Maccary_2026}.} Whereas the first two of these three long GRBs show tentative kilonova associations based on their temporal lags and spectral properties, GRB230307A shows a slightly stronger connection, marked by tellurium emission lines indicative of active $r$-process nucleosynthesis. 

Given the limited understanding of the exact physical processes governing the creation of GRBs, it remains unclear how properties other than the progenitor may determine the classification of GRBs. There are theoretical frameworks that support compact binary merger powered long GRBs~\citep[e.g.,][]{gottlieb2023, rosswog2024mergersdoubleneutronstars}. Thus far, no type of gravitational-wave signal has been confidently observed coincident with a long GRB~\citep{ligo_grb_o3a, ligo_grb_o3b, Wang__2022, Abbott_2019_grb_o2, Aasi_2014, Abbott_2017_grb_o1}. A coincident long GRB–gravitational-wave merger detection would provide unambiguous evidence linking at least some long GRBs to binary mergers. We perform a modelled search for sub-threshold binary neutron star and neutron star--black hole merger gravitational-wave signals in data from the third observing run of the LIGO~\citep{Aasi_2015}, Virgo~\citep{Acernese_2014}, KAGRA~\citep{akutsu2020overview} (LVK) network~\citep{gwtc-3}.\footnote{While \citet{gwtc-3} defines sub-threshold signals as those with an inferred astrophysical compact binary merger probability below 50\%, but a false alarm rate of less than two per day, we use the term more generally to mean signals that are not confidently detected.} We look for mergers coincident with long GRBs detected by Fermi's GRB Monitor~\citep[GBM;][]{meegan_2009}, and Swift's Burst Alert Telescope~\citep[BAT;][]{gehrels_2004, berthelmy_2005, Tohuvavohu_2020}.

The structure of the Paper is as follows. In Sec.~\ref{sec:method}, we outline our method and techniques used to perform the search. We then present the results of our analysis in Sec.~\ref{sec:analysis}, and conclude in Sec.~\ref{sec:conclusion}.

\section{Methodology}
\label{sec:method}

We use a subset of 70 long GRBs detected by Swift/BAT and Fermi/GBM that occurred during the third LVK observing run~\citep{ligo_grb_o3a, ligo_grb_o3b} and had at the two LIGO gravitational-wave observatories running at the time. We obtain gravitational-wave data from the Gravitational‑Wave Open Science Center~\citep{gwosc_o3}. The set of GRBs and the observatories running at the time are listed in Table~\ref{table:GRB_list} in the Appendix. Each of these GRBs have been analysed before by the LVK Collaboration~\citep{ligo_grb_o3a, ligo_grb_o3b}, however those searches were for signals from collapsars, rather than neutron star coalescences. In other words, those LVK searches looked for coherent excess power between gravitational-wave observatories~\citep{Sutton_2010, W_s_2012}, rather than templated searches using binary neutron star and/or neutron star-black hole waveforms (e.g., \citealp{Harry_2011, Williamson_2014, Sathyaprakash_2009}, Sec.~5).

We require LIGO Hanford (H) and LIGO Livingston (L) to be observing for a period of $\unit[{1200}]{s}$ without significant data quality issues around the GRB trigger time. We include Virgo (V) in our analysis if it was observing, but it is not a requirement. Of our 70 long GRBs, there are $48$ with HLV data and $22$ with HL data. There is no additional redshift or progenitor information for these long GRBs~\citep{Finneran_2025}. 

For each long GRB, we perform parameter estimation and calculate the Bayesian evidence for a binary neutron star signal using the \bilby inference library \citep{bilby_paper, bilby_pipe_paper}. We extract the Bayesian evidence from the nested sampling runs for individual detectors, and for a coherent analysis of the LVK network. For details of these calculations, we refer the reader to Sec. V of~\cite{gwtc-3}. For most parameters we employ standard priors from \citet{bilby_pipe_paper}. For the other parameters, we employ priors designed for our long GRB search; see Table~\ref{table:priors}. 

We fix the sky location according to the electromagnetic trigger.
We ignore the sky-location uncertainty from the electromagnetic trigger, which is small compared to the uncertainty in the sky location determined for an equivalent, sub-threshold, gravitational-wave event~\citep{Berlato_2019, berthelmy_2005}. Prompt GRB emission from a binary merger involving at least one neutron star is generally thought to be produced in a short window around the merger time, $t_0$~\citep[e.g.,][]{Zhang_book}. Following~\citet{ligo_grb_o3a, ligo_grb_o3b}, we use the standard window
$t_0=t_{\rm GRB}+[-5,\,1]\,{\rm s}$, where $t_{\rm GRB}$ is the GRB trigger time. 
This assumes the GRB trigger is created as some product of the merger, not from precursor or delayed prompt pulse emission. Our prior for luminosity distance $D_L$ is uniform in comoving volume out to a maximum value of $\unit[2000]{Mpc}$ where we can be sure there is no chance of a gravitational-wave detection.\footnote{The choice of maximum luminosity distance affects the value of the signal evidence. However, we interpret our signal evidence by comparing it with the distribution obtained from off-source data. (In essence, we are calculating a frequentist significance from a Bayesian evidence.) Since the off-source data is analyzed with the same prior, the statistical significance of a candidate event does not depend on the choice of prior boundary.}

We limit the progenitor masses $m_1\in[0.5,\,6]\,M_\odot$ and $m_2\in[0.5,2]\,M_\odot$ with $m_1\ge m_2$. The upper limit on $m_1$ is a conservative assumption that a system with a mass $>6M_\odot$ would not tidally disrupt the neutron star, and therefore not produce a long GRB~\citep{Shibata_2006, Kyutoku_2021}. We do not limit the progenitor spins, $\chi_1$ and $\chi_2$ beyond the assumption the spins are aligned, and use a use prior on the magnitude of $[0,0.99]$.

\begin{table}[h]
\centering
\begin{tabular}{lll}
\hline
\textbf{Parameter}& \textbf{Type} & \textbf{Range} \\
\hline
$m_1$ & uniform & $[0.5 M_\odot, 6 M_\odot] $ \\
$m_2$ & uniform & $[0.5 M_\odot, 2 M_\odot]$ \\
$\chi_1$ & uniform & $[0, 0.99]$ \\
$\chi_2$ & uniform & $[0, 0.99]$ \\
$D_L$  & uniform in comoving volume & $[\unit[50]{Mpc}, \unit[2000]{Mpc}]$\\
RA & fixed & RA$_{\text{GRB}}$ \\
Dec & fixed & DEC$_{\text{GRB}}$  \\
$t_0$ & uniform & $t_{\rm GRB}+[\unit[-5]{s}, \unit[+1]{s}]$ \\
\hline
\end{tabular}
\caption{\label{table:priors}A summary of the priors used in our analysis. Here, $m_i$ and $\chi_i$ are the component masses and dimensionless spins, $D_L$ is the luminosity distance, RA and Dec are the sky location, and $t_0$ is the time of coalescence, where $t_{\rm GRB}$ is the GRB trigger time. For the other binary parameters we employ default priors from~\citet{bilby_pipe_paper}.}
\end{table}

The waveforms are generated using the phenomenological \textsc{IMRPhenomD} model for gravitational-wave signals with non-precessing spin~\citep{imrphenomd_1, imrphenomd_2}. Waveforms with more underlying physics (such as tides or higher-order modes) exist and are used in GWTC-3~\citep{gwtc-3}. However, since we are searching for sub-threshold signals, we opt for fast waveforms, albeit with less precision than state-of-the-art alternatives. 

We use the evidences to calculate the \textit{Bayes coherence ratio}~\citep[BCR;][]{Veitch2010, Isi_2018,bcr,bcr2}, which compares the ``coherent evidence'' (obtained using two or three observatories) to the ``incoherent evidence,'' which does not require consistency between observatories:
\begin{align}
    \text{BCR} = \frac{{\cal Z}_\text{coherent}}{{\cal Z}_\text{incoherent}} ,
\end{align}
where 
\begin{align}
    {\cal Z}_\text{coherent} \equiv \int d\theta \, {\cal L}(d_H, d_L, ... | \theta) \, \pi(\theta) ,
\end{align}
and 
\begin{align}
    {\cal Z}_\text{incoherent} \equiv & \bigg( \int d\theta \, {\cal L}(d_H | \theta) \, \pi(\theta) \bigg) \times\nonumber\\
    & \bigg( \int d\theta' \, {\cal L}(d_L | \theta') \, \pi(\theta') \bigg) ...,
\end{align}
where $d_i$ are the data from the $i$th observatory.

Formally, the BCR is a Bayes factor comparing the hypothesis that there is an astrophysical signal in two or more detectors to the hypothesis that there are uncorrelated noise artefacts in both detectors that look like binary signals.
True signals are consistent between observatories, which lead to large coherent evidences and hence large BCR values.
Noise artefacts, on the other hand, can produce large signal evidences in one detector, but are unlikely to produce consistent signals in two or more observatories, implying $\cal Z_{\rm incoherent}$ is likely to dominate over $\cal Z_{\rm coherent}$.
Thus, astrophysical signals produce large BCR values while noise artefacts produce small ones.
This property has been leveraged to search, e.g., for intermediate mass black holes \citep{avi_bcr}.

\begin{figure*}
    \includegraphics[width=0.9\linewidth]{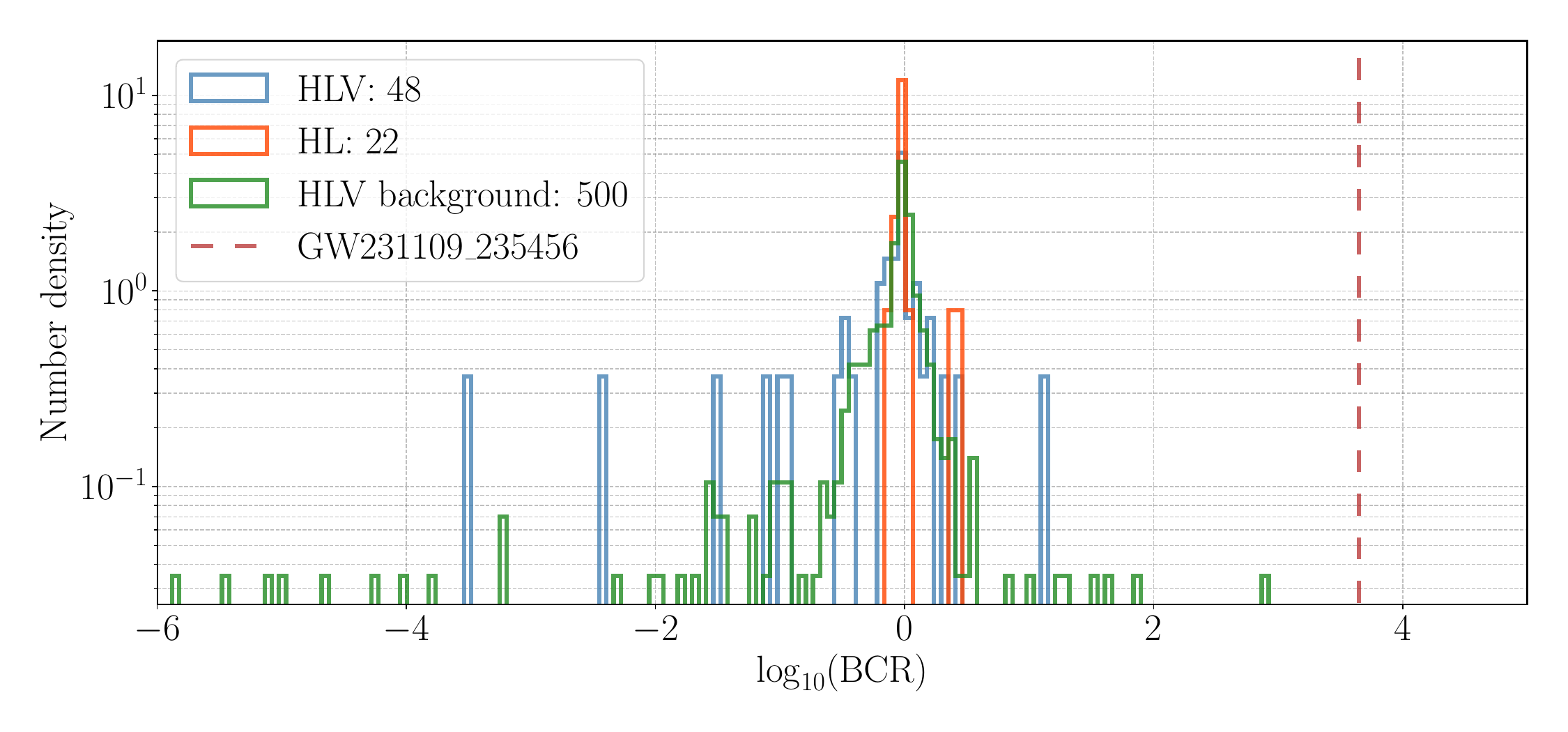}
    \caption{\label{fig:BCR_hist}The distribution of Bayesian coherence ratio (BCR). The distribution for long GRBs coincident with HLV data is shown in blue, the distribution for long GRBs coincident with HL data is shown in orange, and the distribution of off-source data (with no long GRB trigger) is shown in green. The BCR of a recent HL sub-threshold binary neutron-star merger candidate GW231109\_235456~\citep{niu2025gw231109235456subthresholdbinaryneutron} is shown by the maroon vertical line.}
\end{figure*}

\section{Analysis}
\label{sec:analysis}
\subsection{Injection and background studies} \label{sec:test_case}

\begin{figure}
    \includegraphics[width=\linewidth]{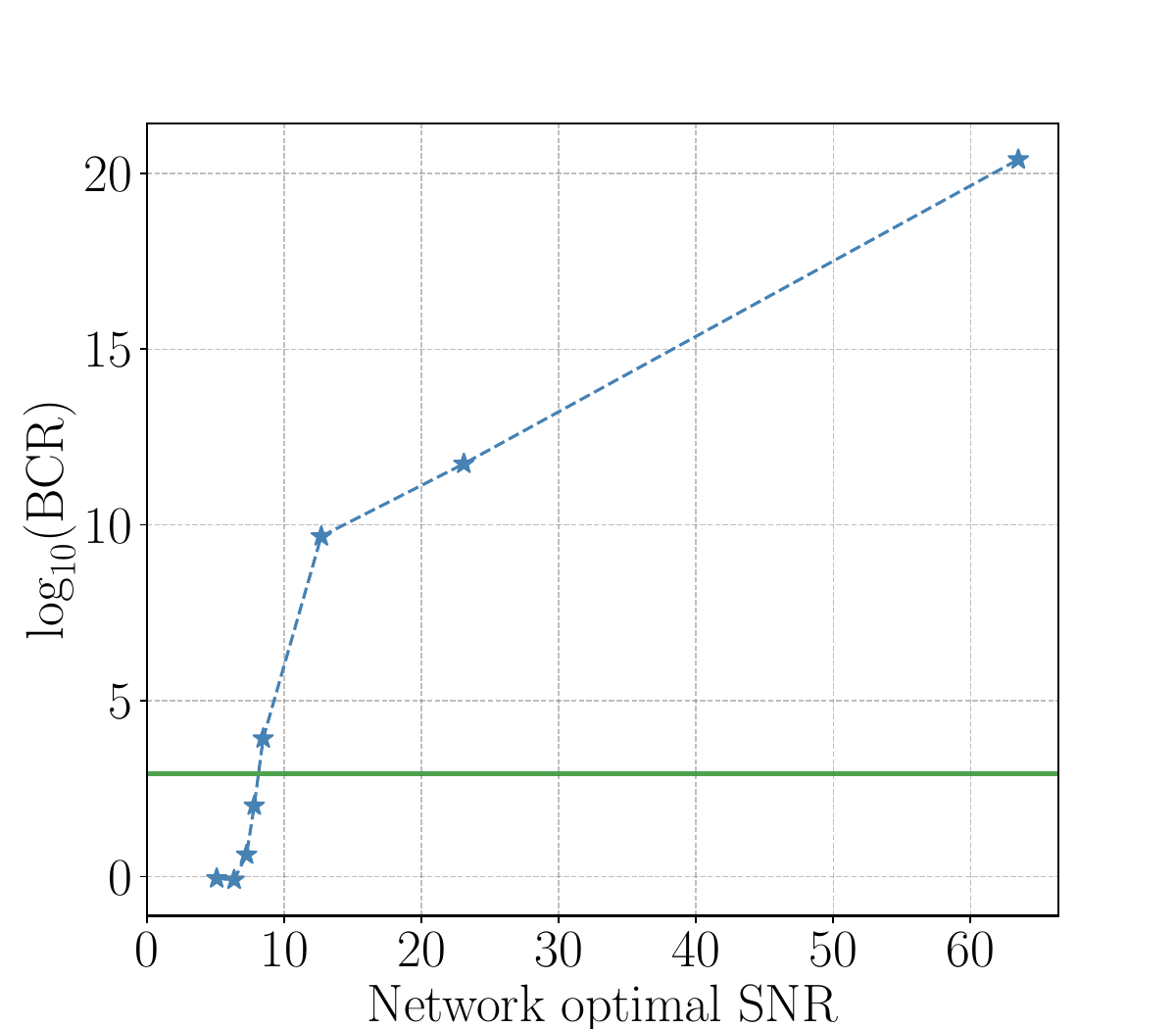}
    \caption{\label{fig:inj_BCRvsSNR}The Bayesian coherence ratio (blue) for a set of injected signals into a three-detector network (HLV) with varying luminosity distance, as described in Section~\ref{sec:test_case}. The background threshold, corresponding to the maximum off-source $\log_{10}(\mathrm{BCR})=2.8$ value, for a positive identification is shown in green. The BCR may provide evidence for sub-threshold events with network optimal SNR $\approx8-12$. 
    }
\end{figure}

We assess the significance of the BCR by performing 500 off-source searches at times around HLV GRB triggers with random sky locations. We discard times with significant data-quality issues. 
In Fig.~\ref{fig:BCR_hist} we plot the distribution of background BCR values in green. 

Based on our background distribution, we determine that a value of $\log_{10}(\text{BCR})>2.8$ is required in order to detect an event with $p=1/500$.
Accounting for trial factors, this corresponds to a $p$-value of 
\begin{align}
    1 - (1-1/500)^{70} = 13\% ,
\end{align}
given that we searched for 70 events.
Thus, we do not expect any events with $\log_{10}(\text{BCR})>2.8$ due to noise fluctuations. Glitches are expected to create a negative tail and there will likely be, by chance, some coherent data across detectors giving some positive values, which are seen in Fig.~\ref{fig:BCR_hist}. 

We verify our method by injecting binary neutron star merger signals into an off-source data segment with a three-detector network (i.e. HLV) and calculating the BCR.
We vary the luminosity distance in order to see how the BCR depends on the network signal-to-noise ratio.
The binary parameters for our injection are summarized in Table~\ref{table:inj_params}.
For the sake of realism, we simulate the signal with a more accurate waveform template, \textsc{IMRPhenomPv2\_NRTidal}~\citep{Dietrich_2019}. 

In Fig.~\ref{fig:inj_BCRvsSNR} we plot $\log_{10}(\text{BCR})$ versus network optimal SNR. The horizontal green line indicates the maximum off-source $\log_{10}(\text{BCR})$ value from our background study. This shows we can identify a signal with a single-event (no trial factors) $p$-value of $0.2\%$ with optimal network SNR $\approx 8.5$ at a luminosity distance of $\unit[300]{Mpc}$.

\begin{table}[h]
\centering
\begin{tabular}{l l}
\hline
\textbf{Parameter} & \textbf{Value} \\
\hline
$m_1$ & $2.1\,M_\odot$ \\
$m_2$ & $1.4\,M_\odot$ \\
$\chi_1$ & $0.01$ \\
$\chi_2$ & $0.01$ \\
$\Lambda_1$ & $600$ \\
$\Lambda_2$ & $600$ \\
$D_L$ & $[40, 110, 200, 300, 325, 350, 400, 500]$\,Mpc \\
RA & $1.215$\,rad \\
Dec & $0.6086$\,rad \\
$\theta_{jn}$ & $0.679$\,rad \\
$\psi$ & $2.424$\,rad \\
$\phi_0$ & $2.198$\,rad \\
$t_0$ & $1260424370.938$\,s \\
\hline
\end{tabular}
\caption{\label{table:inj_params}The parameters of the injected signal. Here, $\chi_1, \chi_2$ are the spin  magnitude of the primary and secondary objects, respectively, $\theta_{jn}$ is the angle between the total angular momentum vector and line of sight, $\psi$ is the polarisation angle, $\phi$ is the binary phase at a given reference frequency, $t_0$ is the geocentric coalescence time, and $\Lambda_1, \Lambda_2$ are the tidal deformabilities of the primary and secondary objects, respectively.}
\end{table}

Additionally, we calculate the BCR of a recent HL sub-threshold binary neutron-star merger candidate GW231109\_235456 \citep{niu2025gw231109235456subthresholdbinaryneutron} and recover a $\log_{10}(\mathrm{BCR})=3.65$, shown on Figures~\ref{fig:BCR_hist} and~\ref{fig:inj_BCRvsSNR} in maroon, above the background thresholds shown in green.
\cite{niu2025gw231109235456subthresholdbinaryneutron} reports a network SNR of 9.7 using their search, which incorporates a redshift-corrected population model. Hypothetically\footnote{Given there is no long GRB trigger, the background study we performed is not the one relevant to this event. Thus, we cannot comment beyond this hypothetical scenario as to the astrophysical nature of this candidate.}, if there had been a corresponding long GRB trigger, our BCR value would suggest the origin is astrophysical, and would be identified as a positive candidate for further investigation.\footnote{We perform a similar validation on GW170817, the first binary neutron star gravitational-wave signal, which had an associated short GRB. This was a three-detector (HLV) event with a network SNR of 32.4. The priors used in our analysis needed to be made stricter in order for the analysis to converge, and gave $\log_{10}(\mathrm{BCR})\approx 10$, broadly consistent with Fig.~\ref{fig:inj_BCRvsSNR}.}

\subsection{Results and implications}\label{sec:fraction}
In Fig.~\ref{fig:BCR_hist}, we show our 48 HLV trigger BCR values in blue and 22 HL triggers in orange. These can be compared to our background BCR values shown in green, which shows that none of our triggers are louder than the largest background value. We conclude that we find no gravitational-wave counterparts. For individual $\text{BCR}$ values for each GRB, see Appendix~\ref{sec:app1}.

We place an upper limit on the fraction of long gamma-ray bursts that are powered by binary mergers. This upper limit is somewhat model-dependent, and so we encourage the reader to treat it as a back-of-the-envelope calculation.

First, we write down the likelihood of observing $N$ gravitationally bright long gamma-ray bursts given that a fraction $f$ of long gamma ray bursts are gravitationally bright.
The likelihood is a Poisson distribution
\begin{align}
    {\cal L}(N | f) = \frac{1}{N!} (f R V t_\text{obs})^N e^{-fRV t_\text{obs}} ,
\end{align}
where $V$ is the sensitive volume of the LVK network, $R$ is the rate density of long gamma-ray bursts (units = $\unit[]{Gpc^{-3}\,yr^{-1}}$), and $t_\text{obs}$ is the duration of the data set.
This likelihood compares the \textit{measured} number of gravitationally bright long GRBs with the \textit{expected} number:
\begin{align}
\hat{N} = f R V t_\text{obs} .
\end{align}

We assume the rate of electromagnetically-bright long GRBs detected by Fermi/Swift is $R = \unit[79^{+57}_{-33}]{Gpc^{-3}\,yr^{-1}}$~\citep{ghirlanda2022cosmichistorylonggamma}. The typical jet opening angle associated with GRBs limits the number of long GRBs detected by Swift/Fermi, and so we apply a beaming correction factor $f_b^{-1}= 260$ (corresponding to a jet opening angle of $\approx 5^\circ$)~\citep{grb_revisited, ghirlanda2007, Liang_2007}.

We observe $N=0$ gravitationally bright long GRBs.
We estimate the sensitive volume using posterior samples from our \bilby runs.
In Sec.~\ref{sec:test_case}, we determine that a signal with network optimal SNR $\sim 9$ would stand out above our background (with $p=17\%$ after trial factors).
For the sake of convenience, we use this same SNR $\sim 9$ to define our sensitive volume.\footnote{This choice is somewhat arbitrary, but other parts of this calculation contribute more systematic uncertainty than any associated with our choice of threshold.}

For each event $i$ we calculate $\hat{d}_i$: the comoving distance---averaged over 10 draws from our prior---with optimal SNR $\rho_{\rm opt}= 9$:
\begin{align}
    \hat{d}_i &= \int d\theta \, d_C(\theta) \, 
    \pi(\theta) \, 
    \Theta(\rho_\text{opt}^i > 9) 
\end{align}
The variable $\hat{d}_i$ is a measure of the ``horizon'' for which we can see binary systems drawn from our prior distribution.
The horizon depends sensitively on our assumptions about the mass distribution of binaries powering long GRBs; more massive binaries can be detected further away.
With that caveat, we calculate the sensitive volume given we are at low redshift: 
\begin{align}
    V \approx \frac{4\pi}{3}\frac{1}{n}
    \sum_{i=1}^n (\hat{d}_i)^3 .
\end{align}

\begin{figure}
    \includegraphics[width=\linewidth]{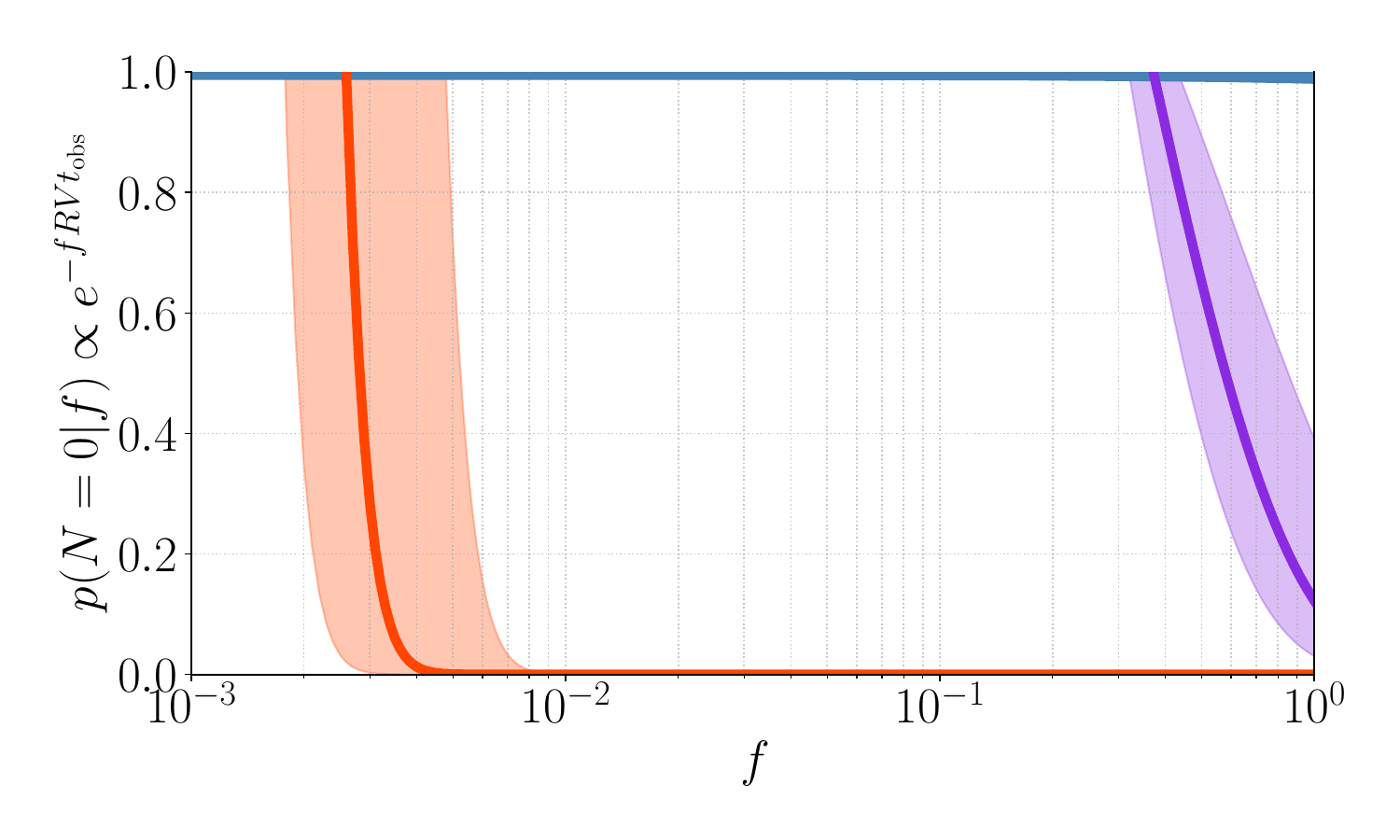}
    \caption{\label{fig:fraction_f}Estimated posterior distribution of $f$, the fraction of gravitationally-bright long GRBs originating from binary mergers. The blue curve shows constraints obtained using the current sensitive search volume, the purple curve assumes a 500-fold increase in sensitive volume, and the orange curve corresponds to a next-generation detector, Cosmic Explorer, under the assumption of no coincident long GRB and gravitational-wave detections, as described in Section~\ref{sec:fraction}. We can then constrain $f$ to an upper limit of 0.90, 0.62, and 0.00 with $90\%$ credibility, respectively. The shaded region plots the uncertainty from the rate of gravitationally-bright GRBs, $R = \unit[79^{+57}_{-33}]{Gpc^{-3}yr^{-1}}$.}
\end{figure}

We now have the ingredients to calculate the posterior for $f$ (the fraction of gravitationally bright long GRBs):
\begin{align}
    p(f | N=0) \propto e^{-fRV} .
\end{align}
The posterior for $f$ is shown in Fig.~\ref{fig:fraction_f}.
The solid blue curve marginalizes over uncertainty in the long GRB rate $R$ while the shaded region shows how uncertainty in $R$ affects our estimate of $f$. Given $RV \approx 0.01$, we find the posterior to be largely flat and uninformative, and predict $f\lesssim0.9$ with 90\% credibility. The solid purple curve shows the same for $RV \approx 4$. This 500-fold increase in sensitive volume provides informative constraints and we find that $f\lesssim0.62$ with 90\% credibility. This may not be achievable with current-generation gravitational-wave detectors, even with further upgrades.

The solid orange curve shows a similar calculation done assuming no associated gravitational waves are observed with the next-generation detector Cosmic Explorer~\citep{evans2021horizonstudycosmicexplorer}, which will be sensitive to binary neutron star mergers up to redshift $\sim 5$ \citep{evans2021horizonstudycosmicexplorer}. Here, we conservatively estimate the sensitive volume as $\unit[1000]{Gpc^{-3}}$, and assume the long GRB rate increases by an order of magnitude between redshift 1 and 3. If thousands of long GRBs lie within the sensitive volume of Cosmic Explorer, $f$ could be measured with high confidence.

Constraints on $f$ under different assumptions remain largely uninformative without a substantial increase in sensitive volume. Even in the optimistic scenario where every successful jet launched by a binary neutron star merger produces a long GRB--corresponding to roughly one-fifth of the predicted merger rate of $\unit[7.6-250]{Gpc^{-3}\,yr^{-1}}$--predicted mergers~\citep{Salafia_2022, LVKgwtc40populationpropertiesmerging})--the expected event rate remains below the observed long-GRB rate. Assuming instead that all binary neutron star and neutron star–black hole mergers produce a successful long GRB only increases the rate by a factor of $\sim 5$. While these estimates are probabilistic, there remains a low but non-zero chance of observing a long GRB within the current sensitive search volume.


\section{Conclusion}
\label{sec:conclusion}
Observations have long supported the bimodal distribution in the GRB population. Core-collapse supernovae have been associated with long GRBs based on sky localisation and spectra~\citep{galama_1998, Hjorth_2003, stanek_2003, hjorth_bloom_2011}. Kilonovae with short GRBs have been associated with compact binary mergers~\citep{Berger_2013,Tanvir_2013,goldstein_2017, savchenko_2017, Abbott_2017_grb, Abbott_2017_170817, Abbott_2019_prop}. To date, the sole coincident GRB and gravitational-wave detection, GW170817 and GRB170817A, confirms binary neutron star mergers are a progenitor of some short GRBs and enabled significant multi-messenger campaigns, opening the door to a new frontier of science~\citep{Abbott_2017_grb, Abbott_2017_multi}. However, this picture has been challenged with evidence linking some long GRBs to binary mergers involving at least one neutron star. 

We aim to fill a gap in the literature created by this new evidence, by performing a modelled search for sub-threshold binary mergers coincident with long GRBs. A confident detection of such an association would fundamentally reshape our understanding of GRB progenitors. We leverage the sky and time localisation provided by long GRBs to improve search sensitivity, and perform Bayesian inference on the data. We find no evidence for gravitational-wave signals coincident with long GRBs and set limits on the luminosity distance to each of these GRBs, assuming a neutron star or neutron star--black hole progenitor, ranging from \unit[50]{Mpc} to \unit[801]{Mpc} (90\% credibility). We test our method on a sub-threshold binary-neutron star merger candidate, GW231109\_235456, with no GRB counterpart, not identified as a candidate via the conventional analysis priors. This method succeeds in identifying the hypothetical candidate with a $\log_{10}(\mathrm{BCR})=3.65$, above the background threshold of $\log_{10}(\mathrm{BCR})=2.80$.

We find that with the current sensitive search volume, we cannot place informative constraints on the fraction of long GRBs with binary merger associations. Using the rate of long GRBs in the local Universe, we estimate that at least a 500-fold increase in the sensitive search volume is required to begin constraining this fraction. There is still a non-zero chance we observe long GRBs within the sensitive volume. Future observing runs with improved detector sensitivity will increase the sensitive volume of our search, while the increased catalog of long GRBs will provide more opportunities for coincident detection. We plan to extend this targeted search approach to data from the fourth observing run of the LVK network, where these combined improvements will either enable the first detection of a gravitational-wave signal from a binary merger coincident with a long GRB, or further constrain the fraction of long GRBs originating from binary mergers involving a neutron star. 

\section{Acknowledgments}
This research is supported by the Australian Research Council Centre of Excellence for Gravitational Wave Discovery (OzGrav), Project Numbers CE170100004 and CE230100016, Discovery Projects DP220101610 and DP230103088, and LIEF Project LE210100002. This material is based upon work supported by NSF's LIGO Laboratory which is a major facility fully funded by the National Science Foundation. The authors are grateful for for computational resources provided by the LIGO Laboratory computing cluster at California Institute of Technology supported by National Science Foundation Grants PHY-0757058 and PHY-0823459. This research has made use of the GRBSN webtool, available at https://grbsn.watchertelescope.ie, and data from the Gravitational Wave Open Science
Center (https://www.gw-openscience.org), a service of
LIGO Laboratory, the LIGO Scientific Collaboration
and the Virgo Collaboration.

\appendix
\section{GRB data}
We summarise the list of long GRBs used in this study in Table.~\ref{table:GRB_list}. These are the GRBs analysed by the LVK Collaboration when searching for collapsar signals~\citep{ligo_grb_o3a,ligo_grb_o3b}. We note the LVK GRB classification is based on the duration over which the fluence of the signal in the relevant GRB detector increases from 5\% to 95\%, denoted $T_{90}$, with associated error, $\delta T_{90}$~\citep{ligo_grb_o3a, ligo_grb_o3b}. There are therefore minor differences in reported $T_{90}$ between detectors (i.e., between Swift and Fermi). The LVK Collaboration sorts GRBs into three categories, 
\[
\mathrm{Class}(T_{90},\delta T_{90}) =
\begin{cases}
\text{short}, & T_{90} + |\delta T_{90}| < 2~\mathrm{s}, \\
\text{long},  & T_{90} - |\delta T_{90}| > 4~\mathrm{s}, \\
\text{ambiguous}, & \text{otherwise}.
\end{cases}
\]
A more stringent classification would include spectral information and burst energetics. 

In Table~\ref{table:GRB_list}, we quote the LVK detector network used for the analysis (HL---LIGO Hanford and LIGO Livingston, HLV---LIGO Hanford, LIGO Livingston, and Virgo), $\log_{10}(\text{BCR})$, and $D_{90}$---the 90\% lower credible bound on $D_L$ posteriors. The $D_L$ posteriors are a measure of the detector sensitivity at that time and sky localisation given we do not observe a merger. Thus, we can set limits on the luminosity distance to each of these long GRBs under the assumption that they were powered by a binary merger ranging from \unit[50]{Mpc} to \unit[801]{Mpc}.

\label{sec:app1}
\begin{table*}[h]
\centering
\begin{tabular}{llcc |lllcc}
\hline
\textbf{GRB Name} & \textbf{Network} & \textbf{$\log_{10}(\text{BCR})$} & \textbf{$D_{90}$\,(Mpc)} 
&& \textbf{GRB Name} & \textbf{Network} & \textbf{$\log_{10}(\text{BCR})$} & \textbf{$D_{90}$\,(Mpc)} \\
\hline
GRB190404293 & HLV & -0.19 & 175 
&& GRB190926A & HL & -0.12 & 294 \\
GRB190406450 & HLV & -3.49 & 51
&& GRB191110A & HLV & -0.47 & 324 \\
GRB190504415 & HLV & -0.02 & 498
&& GRB191111A & HL & -0.04 & 612 \\
GRB190507270 & HL & -0.09 & 187 
&& GRB191119261 & HL & 0.01 & 700 \\
GRB190507970 & HLV & -0.07 & 608 
&& GRB191122A & HLV & -0.07 & 289\\
GRB190511A & HLV & -0.52 & 287
&& GRB191125A & HLV & 1.13 & 273 \\
GRB190512A & HLV & -0.12 & 458 
&& GRB191202A & HLV & -0.42 & 241 \\
GRB190519A & HLV & 0.21 & 394
&& GRB191225B & HL & 0.01 & 683 \\
GRB190603795 & HLV & 0.00 & 702 
&& GRB200105914 & HLV & 0.10 & 150 \\
GRB190604446 & HL & 0.37 & 472 
&& GRB200112395 & HLV & 0.10 & 152 \\
GRB190612165 & HLV & -0.04 & 801
&& GRB200112A & HLV & -0.02 & 639 \\
GRB190613A & HLV & -0.97 & 462 
&& GRB200114A & HLV & 0.01 & 630 \\
GRB190613B & HL & -0.05 & 595 
&& GRB200115A & HL & 0.00 & 492 \\
GRB190615636 & HLV & -9.89 & 50
&& GRB200122A & HLV & 0.01 & 349 \\
GRB190620507 & HL & 0.00 & 671
&& GRB200125B & HL & 0.02 & 467 \\
GRB190628521 & HL & -0.03 & 411 
&& GRB200127B & HL & 0.45 & 169 \\
GRB190630C & HLV & -1.50 & 219
&& GRB200130B & HL & 0.00 & 702 \\
GRB190701A & HLV & -2.40 & 110 
&& GRB200131A & HLV & -0.04 & 371 \\
GRB190707308 & HL & -0.03 & 303 
&& GRB200201A & HLV & -0.02 & 262 \\
GRB190712095 & HLV & -0.20 & 481
&& GRB200205C & HL & -0.09 & 308 \\
GRB190718A & HL & -0.03 & 562
&& GRB200211A & HL & -0.02 & 575\\
GRB190719499 & HLV & -0.05 & 289 
&& GRB200212A & HL & -0.05 & 579 \\
GRB190720613 & HLV & 0.00 & 620
&& GRB200216B & HLV & 0.21 & 184 \\
GRB190720964 & HL & 0.00 & 711 
&& GRB200219A & HLV & 0.31 & 413 \\
GRB190726642 & HLV & 0.02 & 249 
&& GRB200219B & HLV & -0.01 & 374 \\
GRB190726843 & HLV & 0.45 & 322 
&& GRB200224B & HLV & 0.08 & 186\\
GRB190805106 & HLV & -0.21 & 349
&& GRB200227A & HLV & 0.00 & 547 \\
GRB190805199 & HLV & -0.06 & 154 
&& GRB200301320 & HLV & -0.48 & 590\\
GRB190806535 & HLV & -0.16 & 281
&& GRB200303A & HLV & 0.05 & 273 \\
GRB190824A & HL & -0.07 & 593 
&& GRB200308A & HLV & -0.93 & 248 \\
GRB190827467 & HL & 0.00 & 745 
&& GRB200313A & HLV & -0.16 & 418 \\
GRB190831693 & HLV & -1.09 & 183 
&& GRB200317A & HLV & 0.00 & 795 \\
GRB190906767 & HLV & -0.15 & 325
&& GRB200319A & HL & -0.03 & 739 \\
GRB190910028 & HLV & -33.82 & 50 
&& GRB200320A & HLV & -0.01 & 371 \\
GRB190916590 & HLV & 0.13 & 381
&& GRB200326A & HLV & -0.07 & 302 \\
\hline
\end{tabular}
\caption{\label{table:GRB_list}Long GRBs detected by Swift and Fermi that were followed up in our analysis. We report the $\log_{10}(\mathrm{BCR})$ and $D_{90}$---the lower 90\% credible limit on the $D_L$ posterior.}
\end{table*}


\bibliography{refs}

@String{mnras = "Mon. Not. R. Ast. Soc."}

@String{aap = "Astron. Astrophys."}

@String{apj = "Astrophys. J."}

@String{prd = "Phys. Rev. D"}

@String{aa = "Astron. Astrophys."}

@misc{gillanders2025analysisjwstspectrakilonova,
      title={Analysis of the JWST spectra of the kilonova AT 2023vfi accompanying GRB 230307A}, 
      author={J. H. Gillanders and S. J. Smartt},
      year={2025},
      eprint={2408.11093},
      archivePrefix={arXiv},
      primaryClass={astro-ph.HE},
      url={https://arxiv.org/abs/2408.11093}, 
}

@article{Salafia_2022,
   title={Constraints on the merging binary neutron star mass distribution and equation of state based on the incidence of jets in the population},
   volume={666},
   ISSN={1432-0746},
   url={http://dx.doi.org/10.1051/0004-6361/202243260},
   DOI={10.1051/0004-6361/202243260},
   journal={Astronomy \& Astrophysics},
   publisher={EDP Sciences},
   author={Salafia, Om Sharan and Colombo, Alberto and Gabrielli, Francesco and Mandel, Ilya},
   year={2022},
   month=oct, pages={A174} }

@misc{LVKgwtc40populationpropertiesmerging,
      title={GWTC-4.0: Population Properties of Merging Compact Binaries}, 
      author={The LIGO Scientific Collaboration and the Virgo Collaboration and the KAGRA Collaboration, et al.},
      year={2025},
      eprint={2508.18083},
      archivePrefix={arXiv},
      primaryClass={astro-ph.HE},
      url={https://arxiv.org/abs/2508.18083}, 
}

@misc{evans2021horizonstudycosmicexplorer,
      title={A Horizon Study for Cosmic Explorer: Science, Observatories, and Community}, 
      author={Matthew Evans and Rana X Adhikari and Chaitanya Afle and Stefan W. Ballmer and Sylvia Biscoveanu and Ssohrab Borhanian and Duncan A. Brown and Yanbei Chen and Robert Eisenstein and Alexandra Gruson and Anuradha Gupta and Evan D. Hall and Rachael Huxford and Brittany Kamai and Rahul Kashyap and Jeff S. Kissel and Kevin Kuns and Philippe Landry and Amber Lenon and Geoffrey Lovelace and Lee McCuller and Ken K. Y. Ng and Alexander H. Nitz and Jocelyn Read and B. S. Sathyaprakash and David H. Shoemaker and Bram J. J. Slagmolen and Joshua R. Smith and Varun Srivastava and Ling Sun and Salvatore Vitale and Rainer Weiss},
      year={2021},
      eprint={2109.09882},
      archivePrefix={arXiv},
      primaryClass={astro-ph.IM},
      url={https://arxiv.org/abs/2109.09882}, 
}

@article{Liang_2007,
   title={Low‐Luminosity Gamma‐Ray Bursts as a Unique Population: Luminosity Function, Local Rate, and Beaming Factor},
   volume={662},
   ISSN={1538-4357},
   url={http://dx.doi.org/10.1086/517959},
   DOI={10.1086/517959},
   number={2},
   journal={The Astrophysical Journal},
   publisher={American Astronomical Society},
   author={Liang, Enwei and Zhang, Bing and Virgili, Francisco and Dai, Z. G.},
   year={2007},
   month=jun, pages={1111–1118} }

@ARTICLE{ghirlanda2007,
       author = {{Ghirlanda}, G. and {Nava}, L. and {Ghisellini}, G. and {Firmani}, C.},
        title = "{Confirming the {\ensuremath{\gamma}}-ray burst spectral-energy correlations in the era of multiple time breaks}",
      journal = {\aap},
         year = 2007,
        month = apr,
       volume = {466},
       number = {1},
        pages = {127-136},
          doi = {10.1051/0004-6361:20077119},
archivePrefix = {arXiv},
       eprint = {astro-ph/0702352},
 primaryClass = {astro-ph},
       adsurl = {https://ui.adsabs.harvard.edu/abs/2007A&A...466..127G}
}

@ARTICLE{grb_revisited,
       author = {{Wang}, Xiang-Gao and {Zhang}, Bing and {Liang}, En-Wei and {Lu}, Rui-Jing and {Lin}, Da-Bin and {Li}, Jing and {Li}, Long},
        title = "{Gamma-Ray Burst Jet Breaks Revisited}",
      journal = {\apj},
         year = 2018,
        month = jun,
       volume = {859},
       number = {2},
          eid = {160},
        pages = {160},
          doi = {10.3847/1538-4357/aabc13},
archivePrefix = {arXiv},
       eprint = {1804.02113},
 primaryClass = {astro-ph.HE},
       adsurl = {https://ui.adsabs.harvard.edu/abs/2018ApJ...859..160W}
}

@article{bcr2,
author = "G Ashton and E Thrane",
title = "{The astrophysical odds of GW151216}",
journal = mnras,
volume = 498,
pages = "1905",
year = 2020}

@article{bcr,
author = "Gregory Ashton and Eric Thrane and \textbf{R. J. E. Smith}",
title = "{Gravitational wave detection without boot straps: a Bayesian approach}",
journal = prd,
volume = 100,
pages = "123018",
year = 2019}

@misc{ghirlanda2022cosmichistorylonggamma,
      title={The Cosmic History of Long Gamma Ray Bursts}, 
      author={G. Ghirlanda and R. Salvaterra},
      year={2022},
      eprint={2206.06390},
      archivePrefix={arXiv},
      primaryClass={astro-ph.HE},
      url={https://arxiv.org/abs/2206.06390}, 
}

@article{avi_bcr,
author = "A. Vajpeyi and R. Smith and Eric Thrane and others",
title = "{A search for intermediate-mass black holes mergers in the second LIGO--Virgo observing run with the Bayes Coherence Ratio}",
journal = mnras,
volume = "516",
pages = "5309",
year = 2022}

@article{Aasi_2015,
doi = {10.1088/0264-9381/32/7/074001},
url = {https://dx.doi.org/10.1088/0264-9381/32/7/074001},
year = {2015},
month = {mar},
publisher = {IOP Publishing},
volume = {32},
number = {7},
pages = {074001},
author = {J Aasi and others},
title = {{Advanced LIGO}},
journal = {Classical and Quantum Gravity}
}

@article{Acernese_2014,
    author = {Acernese, F and others},
    doi = {10.1088/0264-9381/32/2/024001},
    issn = {1361-6382},
    journal = {Classical and Quantum Gravity},
    month = {December},
    number = {2},
    pages = {024001},
    publisher = {IOP Publishing},
    title = {{Advanced Virgo: a second-generation interferometric gravitational wave detector}},
    url = {http://dx.doi.org/10.1088/0264-9381/32/2/024001},
    volume = {32},
    year = {2014}
}

@misc{akutsu2020overview,
    archiveprefix = {arXiv},
    author = {T. Akutsu and others},
    eprint = {2005.05574},
    primaryclass = {physics.ins-det},
    title = {{Overview of KAGRA: Detector design and construction history}},
    year = {2020}
}

@article{Berlato_2019,
   title={Improved Fermi-GBM GRB Localizations Using BALROG},
   volume={873},
   ISSN={1538-4357},
   url={http://dx.doi.org/10.3847/1538-4357/ab0413},
   DOI={10.3847/1538-4357/ab0413},
   number={1},
   journal={The Astrophysical Journal},
   publisher={American Astronomical Society},
   author={Berlato, F. and Greiner, J. and Burgess, J. Michael},
   year={2019},
   month=mar, pages={60} }

@misc{rosswog2024mergersdoubleneutronstars,
      title={Mergers of double neutron stars with one high-spin component: brighter kilonovae and fallback accretion, weaker gravitational waves}, 
      author={S. Rosswog and P. Diener and F. Torsello and T. M. Tauris and N. Sarin},
      year={2024},
      eprint={2310.15920},
      archivePrefix={arXiv},
      primaryClass={astro-ph.HE},
      url={https://arxiv.org/abs/2310.15920}, 
}

@article{Abbott_2017_grb_o1,
   title={Search for Gravitational Waves Associated with Gamma-Ray Bursts during the First Advanced LIGO Observing Run and Implications for the Origin of GRB 150906B},
   volume={841},
   ISSN={1538-4357},
   url={http://dx.doi.org/10.3847/1538-4357/aa6c47},
   DOI={10.3847/1538-4357/aa6c47},
   number={2},
   journal={The Astrophysical Journal},
   publisher={American Astronomical Society},
   author={Abbott, B. P., et al.},
   year={2017},
   month=may, pages={89} }

@article{Aasi_2014,
   title={Search for Gravitational Waves Associated with gamma-ray Bursts Detected by the Interplanetary Network},
   volume={113},
   ISSN={1079-7114},
   url={http://dx.doi.org/10.1103/PhysRevLett.113.011102},
   DOI={10.1103/physrevlett.113.011102},
   number={1},
   journal={Physical Review Letters},
   publisher={American Physical Society (APS)},
   author={Aasi, J., et al.},
   year={2014},
   month=jun }

@article{Wang__2022,
   title={Search for Coincident Gravitational Waves and Long Gamma-Ray Bursts from 4-OGC and the Fermi-GBM/Swift-BAT Catalog},
   volume={939},
   ISSN={2041-8213},
   url={http://dx.doi.org/10.3847/2041-8213/ac990c},
   DOI={10.3847/2041-8213/ac990c},
   number={1},
   journal={The Astrophysical Journal Letters},
   publisher={American Astronomical Society},
   author={Wang, Yi-Fan and Nitz, Alexander H. and Capano, Collin D. and Wang, Xiangyu Ivy and Yang, Yu-Han and Zhang, Bin-Bin},
   year={2022},
   month=oct, pages={L14} }

@article{gwosc_o3,
   title={Open Data from the Third Observing Run of LIGO, Virgo, KAGRA, and GEO},
   volume={267},
   ISSN={1538-4365},
   url={http://dx.doi.org/10.3847/1538-4365/acdc9f},
   DOI={10.3847/1538-4365/acdc9f},
   number={2},
   journal={The Astrophysical Journal Supplement Series},
   publisher={American Astronomical Society},
   author={Abbott, R. and Abe, H. and Acernese, F. and Ackley, K. and Adhicary, S. and Adhikari, N. and others},
   year={2023},
   month=jul, pages={29} }

@article{Finneran_2025,
   title={The GRBSN webtool: An open-source repository for gamma-ray burst-supernova associations},
   volume={52},
   ISSN={2213-1337},
   url={http://dx.doi.org/10.1016/j.ascom.2025.100954},
   DOI={10.1016/j.ascom.2025.100954},
   journal={Astronomy and Computing},
   publisher={Elsevier BV},
   author={Finneran, Gabriel and Cotter, Laura and Martin-Carrillo, Antonio},
   year={2025},
   month=jul, pages={100954} }

@article{Dietrich_2019,
   title={Improving the NRTidal model for binary neutron star systems},
   volume={100},
   ISSN={2470-0029},
   url={http://dx.doi.org/10.1103/PhysRevD.100.044003},
   DOI={10.1103/physrevd.100.044003},
   number={4},
   journal={Physical Review D},
   publisher={American Physical Society (APS)},
   author={Dietrich, Tim and Samajdar, Anuradha and Khan, Sebastian and Johnson-McDaniel, Nathan K. and Dudi, Reetika and Tichy, Wolfgang},
   year={2019},
   month=aug }

@article{Neights_2025,
   title={GRB 250702B: discovery of a gamma-ray burst from a black hole falling into a star},
   volume={545},
   ISSN={1365-2966},
   url={http://dx.doi.org/10.1093/mnras/staf2019},
   DOI={10.1093/mnras/staf2019},
   number={2},
   journal={Monthly Notices of the Royal Astronomical Society},
   publisher={Oxford University Press (OUP)},
   author={Neights, Eliza and Burns, Eric and Fryer, Chris L and Svinkin, Dmitry and Bala, Suman and Hamburg, Rachel and Gill, Ramandeep and Negro, Michela and Masterson, Megan and DeLaunay, James and Lawrence, David J and Abrahams, Sophie E D and Kawakubo, Yuta and Beniamini, Paz and Diget, Christian Aa and Frederiks, Dmitry and Goldsten, John and Goldstein, Adam and Hall-Smith, Alexander D and Kara, Erin and Laird, Alison M and Lamb, Gavin P and Roberts, Oliver J and Seeb, Ryan and Villar, V Ashley and Airasca, Aldana Holzmann and Barber, Joseph R and Bhat, P Narayana and Bissaldi, Elisabetta and Briggs, Michael S and Cleveland, William H and Dalessi, Sarah and Depalo, Davide and Giles, Misty M and Granot, Jonathan and Hristov, Boyan A and Hui, C Michelle and von Kienlin, Andreas and Kierans, Carolyn and Kocevski, Daniel and Lesage, Stephen and Lysenko, Alexandra L and Mailyan, Bagrat and Malacaria, Christian and Mukherjee, Oindabi and Parsotan, Tyler and Ridnaia, Anna and Ronchini, Samuele and Scotton, Lorenzo and Trigg, Aaron C and Tsvetkova, Anastasia and Ulanov, Mikhail and Veres, Péter and Williams, Maia and Wilson-Hodge, Colleen A and Wood, Joshua},
   year={2025},
   month=nov }

@article{Maccary_2026,
   title={A set of distinctive properties ruling the prompt emission of GRB 230307A and other long γ-ray bursts from compact object mergers},
   volume={49},
   ISSN={2214-4048},
   url={http://dx.doi.org/10.1016/j.jheap.2025.100456},
   DOI={10.1016/j.jheap.2025.100456},
   journal={Journal of High Energy Astrophysics},
   publisher={Elsevier BV},
   author={Maccary, R. and Guidorzi, C. and Maistrello, M. and Kobayashi, S. and Bulla, M. and Moradi, R. and Yi, S.-X. and Wang, C.W. and Zhang, W.L. and Tan, W.-J. and Xiong, S.-L. and Zhang, S.-N.},
   year={2026},
   month=jan, pages={100456} }

@article{Kyutoku_2021,
   title={Coalescence of black hole–neutron star binaries},
   volume={24},
   ISSN={1433-8351},
   url={http://dx.doi.org/10.1007/s41114-021-00033-4},
   DOI={10.1007/s41114-021-00033-4},
   number={1},
   journal={Living Reviews in Relativity},
   publisher={Springer Science and Business Media LLC},
   author={Kyutoku, Koutarou and Shibata, Masaru and Taniguchi, Keisuke},
   year={2021},
   month=dec }

@article{Shibata_2006,
   title={Merger of black hole-neutron star binaries: Nonspinning black hole case},
   volume={74},
   ISSN={1550-2368},
   url={http://dx.doi.org/10.1103/PhysRevD.74.121503},
   DOI={10.1103/physrevd.74.121503},
   number={12},
   journal={Physical Review D},
   publisher={American Physical Society (APS)},
   author={Shibata, Masaru and Uryū, Koji},
   year={2006},
   month=dec }

@article{ligo_grb_o3b,
   title={Search for Gravitational Waves Associated with Gamma-Ray Bursts Detected by Fermi and Swift during the LIGO–Virgo Run O3b},
   volume={928},
   ISSN={1538-4357},
   url={http://dx.doi.org/10.3847/1538-4357/ac532b},
   DOI={10.3847/1538-4357/ac532b},
   number={2},
   journal={The Astrophysical Journal},
   publisher={American Astronomical Society},
   author={Abbott, R. and Abbott, T. D. and Acernese, F. and Ackley, K. and Adams, C. and others},
   year={2022},
   month=apr, pages={186} }

@article{ligo_grb_o3a,
   title={Search for Gravitational Waves Associated with Gamma-Ray Bursts Detected by Fermi and Swift during the LIGO–Virgo Run O3a},
   volume={915},
   ISSN={1538-4357},
   url={http://dx.doi.org/10.3847/1538-4357/abee15},
   DOI={10.3847/1538-4357/abee15},
   number={2},
   journal={The Astrophysical Journal},
   publisher={American Astronomical Society},
   author={Abbott, R. and Abbott, T. D. and Abraham, S. and Acernese, F. and Ackley, K. and Adams, C. and Adhikari, R. X. and Adya, V. B. and others},
   year={2021},
   month=jul, pages={86} }

@article{Sutton_2010,
   title={X-Pipeline: an analysis package for autonomous gravitational-wave burst searches},
   volume={12},
   ISSN={1367-2630},
   url={http://dx.doi.org/10.1088/1367-2630/12/5/053034},
   DOI={10.1088/1367-2630/12/5/053034},
   number={5},
   journal={New Journal of Physics},
   publisher={IOP Publishing},
   author={Sutton, Patrick J and Jones, Gareth and Chatterji, Shourov and Kalmus, Peter and Leonor, Isabel and Poprocki, Stephen and Rollins, Jameson and Searle, Antony and Stein, Leo and Tinto, Massimo and Was, Michal},
   year={2010},
   month=may, pages={053034} }

@article{W_s_2012,
   title={Performance of an externally triggered gravitational-wave burst search},
   volume={86},
   ISSN={1550-2368},
   url={http://dx.doi.org/10.1103/PhysRevD.86.022003},
   DOI={10.1103/physrevd.86.022003},
   number={2},
   journal={Physical Review D},
   publisher={American Physical Society (APS)},
   author={Wąs, Michał and Sutton, Patrick J. and Jones, Gareth and Leonor, Isabel},
   year={2012},
   month=jul }

@article{Williamson_2014,
   title={Improved methods for detecting gravitational waves associated with short gamma-ray bursts},
   volume={90},
   ISSN={1550-2368},
   url={http://dx.doi.org/10.1103/PhysRevD.90.122004},
   DOI={10.1103/physrevd.90.122004},
   number={12},
   journal={Physical Review D},
   publisher={American Physical Society (APS)},
   author={Williamson, A.R. and Biwer, C. and Fairhurst, S. and Harry, I.W. and Macdonald, E. and Macleod, D. and Predoi, V.},
   year={2014},
   month=dec }

@article{Sathyaprakash_2009,
   title={Physics, Astrophysics and Cosmology with Gravitational Waves},
   volume={12},
   ISSN={1433-8351},
   url={http://dx.doi.org/10.12942/lrr-2009-2},
   DOI={10.12942/lrr-2009-2},
   number={1},
   journal={Living Reviews in Relativity},
   publisher={Springer Science and Business Media LLC},
   author={Sathyaprakash, B. S. and Schutz, Bernard F.},
   year={2009},
   month=mar }

@article{Harry_2011,
   title={Targeted coherent search for gravitational waves from compact binary coalescences},
   volume={83},
   ISSN={1550-2368},
   url={http://dx.doi.org/10.1103/PhysRevD.83.084002},
   DOI={10.1103/physrevd.83.084002},
   number={8},
   journal={Physical Review D},
   publisher={American Physical Society (APS)},
   author={Harry, I. W. and Fairhurst, S.},
   year={2011},
   month=apr }

@article{Yang_2022,
   title={A long-duration gamma-ray burst with a peculiar origin},
   volume={612},
   ISSN={1476-4687},
   url={http://dx.doi.org/10.1038/s41586-022-05403-8},
   DOI={10.1038/s41586-022-05403-8},
   number={7939},
   journal={Nature},
   publisher={Springer Science and Business Media LLC},
   author={Yang, Jun and Ai, Shunke and Zhang, Bin-Bin and Zhang, Bing and Liu, Zi-Ke and Wang, Xiangyu Ivy and Yang, Yu-Han and Yin, Yi-Han and Li, Ye and Lü, Hou-Jun},
   year={2022},
   month=dec, pages={232–235} }

@article{Rastinejad_2022,
   title={A kilonova following a long-duration gamma-ray burst at 350 Mpc},
   volume={612},
   ISSN={1476-4687},
   url={http://dx.doi.org/10.1038/s41586-022-05390-w},
   DOI={10.1038/s41586-022-05390-w},
   number={7939},
   journal={Nature},
   publisher={Springer Science and Business Media LLC},
   author={Rastinejad, Jillian C. and Gompertz, Benjamin P. and Levan, Andrew J. and Fong, Wen-fai and Nicholl, Matt and Lamb, Gavin P. and Malesani, Daniele B. and Nugent, Anya E. and Oates, Samantha R. and Tanvir, Nial R. and de Ugarte Postigo, Antonio and Kilpatrick, Charles D. and Moore, Christopher J. and Metzger, Brian D. and Ravasio, Maria Edvige and Rossi, Andrea and Schroeder, Genevieve and Jencson, Jacob and Sand, David J. and Smith, Nathan and Fernández, José Feliciano Agüí and Berger, Edo and Blanchard, Peter K. and Chornock, Ryan and Cobb, Bethany E. and De Pasquale, Massimiliano and Fynbo, Johan P. U. and Izzo, Luca and Kann, D. Alexander and Laskar, Tanmoy and Marini, Ester and Paterson, Kerry and Escorial, Alicia Rouco and Sears, Huei M. and Thöne, Christina C.},
   year={2022},
   month=dec, pages={223–227} }

@article{Abbott_2017_grb,
   title={Gravitational Waves and Gamma-Rays from a Binary Neutron Star Merger: GW170817 and GRB 170817A},
   volume={848},
   ISSN={2041-8213},
   url={http://dx.doi.org/10.3847/2041-8213/aa920c},
   DOI={10.3847/2041-8213/aa920c},
   number={2},
   journal={The Astrophysical Journal Letters},
   publisher={American Astronomical Society},
   author={Abbott, B. P. and Abbott, R. and Abbott, T. D. and Acernese, F. and Ackley, K. and Adams, C. and Adams, T. and Addesso, P. and Adhikari, R. X. and others},
   year={2017},
   month=oct, pages={L13} }

@misc{hjorth_bloom_2011,
      title={The Gamma-Ray Burst - Supernova Connection}, 
      author={Jens Hjorth and Joshua S. Bloom},
      year={2011},
      eprint={1104.2274},
      archivePrefix={arXiv},
      primaryClass={astro-ph.HE},
      url={https://arxiv.org/abs/1104.2274}, 
}

@article{Berger_2013,
   title={AN r-PROCESS KILONOVA ASSOCIATED WITH THE SHORT-HARD GRB 130603B},
   volume={774},
   ISSN={2041-8213},
   url={http://dx.doi.org/10.1088/2041-8205/774/2/L23},
   DOI={10.1088/2041-8205/774/2/l23},
   number={2},
   journal={The Astrophysical Journal},
   publisher={American Astronomical Society},
   author={Berger, E. and Fong, W. and Chornock, R.},
   year={2013},
   month=aug, pages={L23} }

@article{Tohuvavohu_2020,
   title={Gamma-Ray Urgent Archiver for Novel Opportunities (GUANO): Swift/BAT Event Data Dumps on Demand to Enable Sensitive Subthreshold GRB Searches},
   volume={900},
   ISSN={1538-4357},
   url={http://dx.doi.org/10.3847/1538-4357/aba94f},
   DOI={10.3847/1538-4357/aba94f},
   number={1},
   journal={The Astrophysical Journal},
   publisher={American Astronomical Society},
   author={Tohuvavohu, Aaron and Kennea, Jamie A. and DeLaunay, James and Palmer, David M. and Cenko, S. Bradley and Barthelmy, Scott},
   year={2020},
   month=aug, pages={35} }

@ARTICLE{berthelmy_2005,
       author = {{Barthelmy}, Scott D. and {Barbier}, Louis M. and {Cummings}, Jay R. and {Fenimore}, Ed E. and {Gehrels}, Neil and {Hullinger}, Derek and {Krimm}, Hans A. and {Markwardt}, Craig B. and {Palmer}, David M. and {Parsons}, Ann and {Sato}, Goro and {Suzuki}, Masaya and {Takahashi}, Tadayuki and {Tashiro}, Makota and {Tueller}, Jack},
        title = "{The Burst Alert Telescope (BAT) on the SWIFT Midex Mission}",
      journal = {\ssr},
         year = 2005,
        month = oct,
       volume = {120},
       number = {3-4},
        pages = {143-164},
          doi = {10.1007/s11214-005-5096-3},
archivePrefix = {arXiv},
       eprint = {astro-ph/0507410},
 primaryClass = {astro-ph},
       adsurl = {https://ui.adsabs.harvard.edu/abs/2005SSRv..120..143B}
}

@ARTICLE{gehrels_2004,
       author = {{Gehrels}, N. and {Chincarini}, G. and {Giommi}, P. and {Mason}, K.~O. and {Nousek}, J.~A. and {Wells}, A.~A. and {White}, N.~E. and {Barthelmy}, S.~D. and {Burrows}, D.~N. and {Cominsky}, L.~R. and {Hurley}, K.~C. and {Marshall}, F.~E. and {M{\'e}sz{\'a}ros}, P. and {Roming}, P.~W.~A. and {Angelini}, L. and {Barbier}, L.~M. and {Belloni}, T. and {Campana}, S. and {Caraveo}, P.~A. and {Chester}, M.~M. and {Citterio}, O. and {Cline}, T.~L. and {Cropper}, M.~S. and {Cummings}, J.~R. and {Dean}, A.~J. and {Feigelson}, E.~D. and {Fenimore}, E.~E. and {Frail}, D.~A. and {Fruchter}, A.~S. and {Garmire}, G.~P. and {Gendreau}, K. and {Ghisellini}, G. and {Greiner}, J. and {Hill}, J.~E. and {Hunsberger}, S.~D. and {Krimm}, H.~A. and {Kulkarni}, S.~R. and {Kumar}, P. and {Lebrun}, F. and {Lloyd-Ronning}, N.~M. and {Markwardt}, C.~B. and {Mattson}, B.~J. and {Mushotzky}, R.~F. and {Norris}, J.~P. and {Osborne}, J. and {Paczynski}, B. and {Palmer}, D.~M. and {Park}, H. -S. and {Parsons}, A.~M. and {Paul}, J. and {Rees}, M.~J. and {Reynolds}, C.~S. and {Rhoads}, J.~E. and {Sasseen}, T.~P. and {Schaefer}, B.~E. and {Short}, A.~T. and {Smale}, A.~P. and {Smith}, I.~A. and {Stella}, L. and {Tagliaferri}, G. and {Takahashi}, T. and {Tashiro}, M. and {Townsley}, L.~K. and {Tueller}, J. and {Turner}, M.~J.~L. and {Vietri}, M. and {Voges}, W. and {Ward}, M.~J. and {Willingale}, R. and {Zerbi}, F.~M. and {Zhang}, W.~W.},
        title = "{The Swift Gamma-Ray Burst Mission}",
      journal = {\apj},
         year = 2004,
        month = aug,
       volume = {611},
       number = {2},
        pages = {1005-1020},
          doi = {10.1086/422091},
archivePrefix = {arXiv},
       eprint = {astro-ph/0405233},
 primaryClass = {astro-ph},
       adsurl = {https://ui.adsabs.harvard.edu/abs/2004ApJ...611.1005G}
}

@ARTICLE{meegan_2009,
       author = {{Meegan}, Charles and {Lichti}, Giselher and {Bhat}, P.~N. and {Bissaldi}, Elisabetta and {Briggs}, Michael S. and {Connaughton}, Valerie and {Diehl}, Roland and {Fishman}, Gerald and {Greiner}, Jochen and {Hoover}, Andrew S. and {van der Horst}, Alexander J. and {von Kienlin}, Andreas and {Kippen}, R. Marc and {Kouveliotou}, Chryssa and {McBreen}, Sheila and {Paciesas}, W.~S. and {Preece}, Robert and {Steinle}, Helmut and {Wallace}, Mark S. and {Wilson}, Robert B. and {Wilson-Hodge}, Colleen},
        title = "{The Fermi Gamma-ray Burst Monitor}",
      journal = {\apj},
         year = 2009,
        month = sep,
       volume = {702},
       number = {1},
        pages = {791-804},
          doi = {10.1088/0004-637X/702/1/791},
archivePrefix = {arXiv},
       eprint = {0908.0450},
 primaryClass = {astro-ph.IM},
       adsurl = {https://ui.adsabs.harvard.edu/abs/2009ApJ...702..791M}
}

@misc{zhu2025,
      title={Identifying Merger-Driven Long Gamma-Ray Bursts based on Machine Learning}, 
      author={Si-Yuan Zhu and Hui-Ying Deng and Fu-Wen Zhang and Qian-Zi Mo and Pak-Hin Thomas Tam},
      year={2025},
      eprint={2506.08675},
      archivePrefix={arXiv},
      primaryClass={astro-ph.HE},
      url={https://arxiv.org/abs/2506.08675}, 
}

@misc{gottlieb2023,
      title={A Unified Picture of Short and Long Gamma-ray Bursts from Compact Binary Mergers}, 
      author={Ore Gottlieb and Brian Metzger and Eliot Quataert and Danat Issa and Tia Martineau and Francois Foucart and Matthew Duez and Lawrence Kidder and Harald Pfeiffer and Mark Scheel},
      year={2023},
      eprint={2309.00038},
      archivePrefix={arXiv},
      primaryClass={astro-ph.HE},
      url={https://arxiv.org/abs/2309.00038}, 
}

@article{Jin_2015,
   title={THE LIGHT CURVE OF THE MACRONOVA ASSOCIATED WITH THE LONG–SHORT BURST GRB 060614},
   volume={811},
   ISSN={2041-8213},
   url={http://dx.doi.org/10.1088/2041-8205/811/2/L22},
   DOI={10.1088/2041-8205/811/2/l22},
   number={2},
   journal={The Astrophysical Journal},
   publisher={American Astronomical Society},
   author={Jin, Zhi-Ping and Li, Xiang and Cano, Zach and Covino, Stefano and Fan, Yi-Zhong and Wei, Da-Ming},
   year={2015},
   month=sep, pages={L22} }

@article{Zhang_2007,
   title={Making a Short Gamma-Ray Burst from a Long One: Implications for the Nature of GRB 060614},
   volume={655},
   ISSN={1538-4357},
   url={http://dx.doi.org/10.1086/511781},
   DOI={10.1086/511781},
   number={1},
   journal={The Astrophysical Journal},
   publisher={American Astronomical Society},
   author={Zhang, Bing and Zhang, Bin-Bin and Liang, En-Wei and Gehrels, Neil and Burrows, David N. and Mészáros, Peter},
   year={2007},
   month=jan, pages={L25–L28} }

@ARTICLE{della_2006,
       author = {{Della Valle}, M. and {Chincarini}, G. and {Panagia}, N. and {Tagliaferri}, G. and {Malesani}, D. and {Testa}, V. and {Fugazza}, D. and {Campana}, S. and {Covino}, S. and {Mangano}, V. and {Antonelli}, L.~A. and {D'Avanzo}, P. and {Hurley}, K. and {Mirabel}, I.~F. and {Pellizza}, L.~J. and {Piranomonte}, S. and {Stella}, L.},
        title = "{An enigmatic long-lasting {\ensuremath{\gamma}}-ray burst not accompanied by a bright supernova}",
      journal = {\nat},
         year = 2006,
        month = dec,
       volume = {444},
       number = {7122},
        pages = {1050-1052},
          doi = {10.1038/nature05374},
archivePrefix = {arXiv},
       eprint = {astro-ph/0608322},
 primaryClass = {astro-ph},
       adsurl = {https://ui.adsabs.harvard.edu/abs/2006Natur.444.1050D}
}

@ARTICLE{gehrel_2006,
       author = {{Gehrels}, N. and {Norris}, J.~P. and {Barthelmy}, S.~D. and {Granot}, J. and {Kaneko}, Y. and {Kouveliotou}, C. and {Markwardt}, C.~B. and {M{\'e}sz{\'a}ros}, P. and {Nakar}, E. and {Nousek}, J.~A. and {O'Brien}, P.~T. and {Page}, M. and {Palmer}, D.~M. and {Parsons}, A.~M. and {Roming}, P.~W.~A. and {Sakamoto}, T. and {Sarazin}, C.~L. and {Schady}, P. and {Stamatikos}, M. and {Woosley}, S.~E.},
        title = "{A new {\ensuremath{\gamma}}-ray burst classification scheme from GRB060614}",
      journal = {\nat},
         year = 2006,
        month = dec,
       volume = {444},
       number = {7122},
        pages = {1044-1046},
          doi = {10.1038/nature05376},
archivePrefix = {arXiv},
       eprint = {astro-ph/0610635},
 primaryClass = {astro-ph},
       adsurl = {https://ui.adsabs.harvard.edu/abs/2006Natur.444.1044G}
}

@misc{yi2025,
      title={Long Pulse by Short Central Engine: Prompt emission from expanding dissipation rings in the jet front of gamma-ray bursts}, 
      author={Shu-Xu Yi and Emre Seyit Yorgancioglu and S. -L. Xiong and S. -N. Zhang},
      year={2025},
      eprint={2411.16174},
      archivePrefix={arXiv},
      primaryClass={astro-ph.HE},
      url={https://arxiv.org/abs/2411.16174}, 
}

@article{Popham_1999,
   title={Hyperaccreting Black Holes and Gamma‐Ray Bursts},
   volume={518},
   ISSN={1538-4357},
   url={http://dx.doi.org/10.1086/307259},
   DOI={10.1086/307259},
   number={1},
   journal={The Astrophysical Journal},
   publisher={American Astronomical Society},
   author={Popham, Robert and Woosley, S. E. and Fryer, Chris},
   year={1999},
   month=jun, pages={356–374} }

@article{Paczy_ski_1998,
   title={Are Gamma-Ray Bursts in Star-Forming Regions?},
   volume={494},
   ISSN={0004-637X},
   url={http://dx.doi.org/10.1086/311148},
   DOI={10.1086/311148},
   number={1},
   journal={The Astrophysical Journal},
   publisher={American Astronomical Society},
   author={Paczyński, Bohdan},
   year={1998},
   month=feb, pages={L45–L48} }

@article{Narayan_1992,
   title={Gamma-ray bursts as the death throes of massive binary stars},
   volume={395},
   ISSN={1538-4357},
   url={http://dx.doi.org/10.1086/186493},
   DOI={10.1086/186493},
   journal={The Astrophysical Journal},
   publisher={American Astronomical Society},
   author={Narayan, Ramesh and Paczynski, Bohdan and Piran, Tsvi},
   year={1992},
   month=aug, pages={L83} }

@misc{yang_2023,
      title={A lanthanide-rich kilonova in the aftermath of a long gamma-ray burst}, 
      author={Yu-Han Yang and Eleonora Troja and Brendan O'Connor and Chris L. Fryer and Myungshin Im and Joe Durbak and Gregory S. H. Paek and Roberto Ricci and Clécio R. De Bom and James H. Gillanders and Alberto J. Castro-Tirado and Zong-Kai Peng and Simone Dichiara and Geoffrey Ryan and Hendrik van Eerten and Zi-Gao Dai and Seo-Won Chang and Hyeonho Choi and Kishalay De and Youdong Hu and Charles D. Kilpatrick and Alexander Kutyrev and Mankeun Jeong and Chung-Uk Lee and Martin Makler and Felipe Navarete and Ignacio Pérez-García},
      year={2023},
      eprint={2308.00638},
      archivePrefix={arXiv},
      primaryClass={astro-ph.HE},
      url={https://arxiv.org/abs/2308.00638}, 
}

@misc{sun_2024,
      title={Magnetar emergence in a peculiar gamma-ray burst from a compact star merger}, 
      author={H. Sun and C. -W. Wang and J. Yang and B. -B. Zhang and S. -L. Xiong and Y. -H. I. Yin and Y. Liu and Y. Li and W. -C. Xue and Z. Yan and C. Zhang and W. -J. Tan and H. -W. Pan and J. -C. Liu and H. -Q. Cheng and Y. -Q. Zhang and J. -W. Hu and C. Zheng and Z. -H. An and C. Cai and Z. -M. Cai and L. Hu and C. Jin and D. -Y. Li and X. -Q. Li and H. -Y. Liu and M. Liu and W. -X. Peng and L. -M. Song and S. -L. Sun and X. -J. Sun and X. -L. Wang and X. -Y. Wen and S. Xiao and S. -X. Yi and F. Zhang and W. -D. Zhang and X. -F. Zhang and Y. -H. Zhang and D. -H. Zhao and S. -J. Zheng and Z. -X. Ling and S. -N. Zhang and W. Yuan and B. Zhang},
      year={2024},
      eprint={2307.05689},
      archivePrefix={arXiv},
      primaryClass={astro-ph.HE},
      url={https://arxiv.org/abs/2307.05689}, 
}

@article{Levan_2023,
   title={Heavy-element production in a compact object merger observed by JWST},
   volume={626},
   ISSN={1476-4687},
   url={http://dx.doi.org/10.1038/s41586-023-06759-1},
   DOI={10.1038/s41586-023-06759-1},
   number={8000},
   journal={Nature},
   publisher={Springer Science and Business Media LLC},
   author={Levan, Andrew J. and Gompertz, Benjamin P. and Salafia, Om Sharan and Bulla, Mattia and Burns, Eric and Hotokezaka, Kenta and Izzo, Luca and Lamb, Gavin P. and Malesani, Daniele B. and Oates, Samantha R. and Ravasio, Maria Edvige and Rouco Escorial, Alicia and Schneider, Benjamin and Sarin, Nikhil and Schulze, Steve and Tanvir, Nial R. and Ackley, Kendall and Anderson, Gemma and Brammer, Gabriel B. and Christensen, Lise and Dhillon, Vikram S. and Evans, Phil A. and Fausnaugh, Michael and Fong, Wen-fai and Fruchter, Andrew S. and Fryer, Chris and Fynbo, Johan P. U. and Gaspari, Nicola and Heintz, Kasper E. and Hjorth, Jens and Kennea, Jamie A. and Kennedy, Mark R. and Laskar, Tanmoy and Leloudas, Giorgos and Mandel, Ilya and Martin-Carrillo, Antonio and Metzger, Brian D. and Nicholl, Matt and Nugent, Anya and Palmerio, Jesse T. and Pugliese, Giovanna and Rastinejad, Jillian and Rhodes, Lauren and Rossi, Andrea and Saccardi, Andrea and Smartt, Stephen J. and Stevance, Heloise F. and Tohuvavohu, Aaron and van der Horst, Alexander and Vergani, Susanna D. and Watson, Darach and Barclay, Thomas and Bhirombhakdi, Kornpob and Breedt, Elmé and Breeveld, Alice A. and Brown, Alexander J. and Campana, Sergio and Chrimes, Ashley A. and D’Avanzo, Paolo and D’Elia, Valerio and De Pasquale, Massimiliano and Dyer, Martin J. and Galloway, Duncan K. and Garbutt, James A. and Green, Matthew J. and Hartmann, Dieter H. and Jakobsson, Páll and Kerry, Paul and Kouveliotou, Chryssa and Langeroodi, Danial and Le Floc’h, Emeric and Leung, James K. and Littlefair, Stuart P. and Munday, James and O’Brien, Paul and Parsons, Steven G. and Pelisoli, Ingrid and Sahman, David I. and Salvaterra, Ruben and Sbarufatti, Boris and Steeghs, Danny and Tagliaferri, Gianpiero and Thöne, Christina C. and de Ugarte Postigo, Antonio and Kann, David Alexander},
   year={2023},
   month=oct, pages={737–741} }

@ARTICLE{Paczynski_1991,
       author = {{Paczynski}, Bohdan},
        title = "{Cosmological gamma-ray bursts.}",
      journal = {\actaa},
         year = 1991,
        month = jan,
       volume = {41},
        pages = {257-267},
       adsurl = {https://ui.adsabs.harvard.edu/abs/1991AcA....41..257P}
}
\end{document}